\documentstyle[prl,aps,floats,graphicx]{revtex}
\begin{document}
\font\bfield=cmbxti12
\font\smalsymb=cmtt10 scaled 1000
\draft

\title{Heating of Two-Dimensional Holes in SiGe and the {\bfield B} = 0 
Metal-Insulator Transition}

\author{R. Leturcq, D. L'Hote, R. Tourbot}

\address{Service de Physique de l'Etat Condens\'{e}, 
CEA, Centre d'Etudes de Saclay, F-91191 
Gif-sur-Yvette, France.}

\author{V. Senz$^1$, U. Gennser$^2$, T. Ihn$^1$, K. Ensslin$^1$, 
G. Dehlinger$^3$, D. Gr{\"u}tzmacher$^3$}

\address{$^1$ Laboratory of Solid State Physics, 
ETH Z{\"u}rich, CH-8093 Z{\"u}rich, Switzerland.\\
$^2$ Laboratoire de Photonique et de Nanostructures, CNRS, 92222 Bagneux, France.\\
$^3$ Paul Scherrer Institut, CH-5232 Villigen PSI, Switzerland.}

\date{\today }

\twocolumn[
	\begin{@twocolumnfalse}
		\maketitle
		\widetext
		\begin{abstract}
We study the 
resistivity vs. electric field
dependence $\rho(E)$ of a 2D hole system in SiGe
close to the $B=0$ metal-insulator transition.
Using $\rho$ as a ``thermometer'' to obtain the 
effective temperature of the holes $T_e(E)$, we find that 
the $\rho(E)$ dependence can be attributed to hole heating. 
The hole-phonon coupling
involves weakly screened piezoelectric and deformation
potentials compatible
with previous measurements. The damping of the Shubnikov-de Haas
oscillations gives the same $T_e$ values.
Thus the $\rho(E)$ dependence and the $E$-field ``scaling'' do not 
provide additional evidence for
a quantum phase transition (QPT). We discuss how to
study, in general, true $E$-field scaling and extract the ratio of the QPT
characteristic lengths.
\end{abstract}
\pacs{PACS numbers: 71.30.+h, 63.20.Kr, 73.43.Nq}
\narrowtext
\end{@twocolumnfalse}
]

The recent observations of a metal-insulator transition (MIT)
in two-dimensional (2D) electron or hole systems in high
mobility silicon metal-oxide-semiconductor field effect
transistors (Si-MOSFETs) \cite{Kra1,Pop1} and in certain
heterostructures \cite{Han1,Sim1,Lam1,Col1,Sen1,Abr1} have triggered 
important experimental and theoretical efforts to
understand the unexpected metallic behavior \cite{Abr1,Alt1}. 
This consists in a
decrease of the resistivity $\rho$ for decreasing temperature $T$, 
at densities $p_s$ larger than the critical 
density $p_c$. In contrast, $d\rho/dT<0$ on the insulating 
side of the MIT ($p_s<p_c$). 
The extrapolated finite resistivity at $T$=0 
is in conflict with the scaling theory of localization for non-interacting 
particles
which predicts localized states in 2D systems\cite{Abr2}.
The MIT occurs for values of $r_s$ 
(the ratio of a carrier pair Coulomb energy to the
Fermi energy) much larger than one, suggesting the existence 
of a non-Fermi liquid due to strong interactions. The MIT could thus
be the signature of a $T$=0 quantum phase transition (QPT) between new
metallic and insulating phases.
However, for some systems, e.g. $p$-SiGe, it seems that
the positive $d\rho/dT$ is due to quasi-classical processes masking the 
usual weak
localization which should appear again at low enough temperatures 
\cite{Alt1,Pud1,Sen2,Sim2}. At present, the
nature of the MIT 
is still a subject of ongoing discussions.

A striking feature of the MIT is that at low temperature, 
the dependence of $\rho$ as a function of the electric field $E$
is similar to $\rho(T)$, i.e.
$d\rho/dE < 0$ for $p_s<p_c$ and $d\rho/dE > 0$ for $p_s>p_c$ 
\cite{Sim1,Sen1,Yoo1,Kra2,Hee1,Lil1}. The physical 
origin of this observation is an open question. Irrespective of the 
possible microscopic 
explanations of the $\rho(E,p_s)$ dependence\cite{Yoo1,Lea1},
a QPT implies that $\rho$ scales with $E$ if it scales with $T$
\cite{Son1}: close to the critical point, two characteristic
length scales are associated with $T$ and $E$,
$L_{\Phi}(T)\sim T^{-1/z}$ and $L_E(E)\sim E^{-1/(z+1)}$, and
$\rho$ depends only on the
ratio of the smallest of them to 
the correlation length $\xi\sim \left| \delta_n \right| ^{-\nu}$ 
($z$ is the dynamical exponent, 
$\nu$ the correlation length exponent, and
$ \delta_n  =(p_s-p_c)/p_c$). 
Thus $\rho(T,p_s)$ (for 
$E\rightarrow0$) depends only on 
$\left| \delta_n \right| /T^{1/z\nu}$ on each side 
of the transition, and $\rho(E,p_s)$ (at low 
temperature) depends only on 
$\left| \delta_n \right| /E^{1/(z+1)\nu}$. 
The scaling analysis of 
$\rho(T,p_s)$ and $\rho(E,p_s)$ allows a separate extraction of $z$ and $\nu$.
$E$ scaling has been observed 
in several systems exhibiting $T$ 
scaling\ \cite{Sen1,Kra2,Hee1,Lil1}.
However, a natural question 
is whether the $\rho(E)$ dependence results
from carrier heating\cite{Son1}.
At low temperature, the weak carrier-phonon coupling can lead to
an effective
temperature of the carriers, $T_e(E)$, larger than the lattice 
temperature $T_l$ 
\cite{Alt1,Pri1,Xie1,All1,Cho1,Zie1,Fle1,Ans1,Fle2,Ken1}. Thus, the
$\rho(E)$ dependence can be due to the 
$T_e(E)$ dependence of $\rho$.

In this paper, we demonstrate experimentally that for a 2D hole 
system in $p$-SiGe exhibiting the MIT features with $T$ and $E$ scaling, 
the resistivity vs. $E$-field dependence can 
be attributed to hole heating. 
We find $\rho(E,p_s)=\rho(T=T_e(E),p_s)$ with 
a law $T_e(E)$ compatible both in shape and magnitude with known 
electron-phonon coupling models and data.
The same $T_e(E)$ is obtained
from the damping of the Shubnikov-de Haas (SdH) oscillations. 
Thus, in the present case, the $\rho(E,p_s)$ dependencies and
the $E$-field ``scaling'' are not arguments 
in favor of the QPT interpretation of the ``MIT''
and the separate extraction of $\nu$ 
and $z$ is not possible. 
As heating effects depend on the semiconductor structure,
while $E$ scaling is a crucial issue for the QPT,
we discuss the conditions required to study it for various systems
in spite of carrier heating. 
We show how the ratio
$L_{\Phi}/L_E$ can be obtained experimentally.

The experiments were performed on 
Si$_{0.85}$Ge$_{0.15}$ quantum wells sandwiched 
between undoped Si layers \cite{Sen1,Sen2}. 
The 2D hole gas was formed in the 
triangular potential well at the 
Si/SiGe interface located on the boron 
doped side. It occupies the 
lowest heavy-hole subband, with an effective mass of about $0.25m_0$.
Two gated Hall bars (S1 and S2) were used. Their width $W$ is 100 $\mu$m, 
and their length $L$ (between voltage probes) 
233 $\mu$m for S1, and 125 $\mu$m for S2. By varying the gate voltage, 
their densities can be tuned between 0.50 and 
$1.75\times10^{11}$ cm$^{-2}$. 
The mobilities at 200 mK increase with $p_s$ 
from 600 to 5500 cm$^2 /$Vs (resp. 1000 to 7400 cm$^2 /$Vs) 
for S1 (resp. S2). 
An ungated sample, 
with a density of $3.9\times10^{11}$ cm$^{-2}$ and a mobility of 
7800 cm$^2 /$Vs has also been used. 
The temperature range was
$70$ mK - 1.4 K.
To check that the temperature $T$ given by the thermometer
fixed in the copper sample holder was that of the 
lattice $T_l$, the low current resistance of another Hall bar 
etched onto the same substrate was used as a thermometer. Its temperature 
remained close to $T$ whatever the current in our samples. 
$\rho=(V/I)(W/L)$ and $E=V/L$ were obtained
from the current $I$ and the voltage drop $V$ between the 
voltage probes, using a four point DC technique, with a 
current (10 - 300 pA 
for low $E$ measurements) periodically reversed at a 
frequency 0.03 - 0.3 Hz. 

\begin{figure}
\includegraphics*[width=8.6cm]{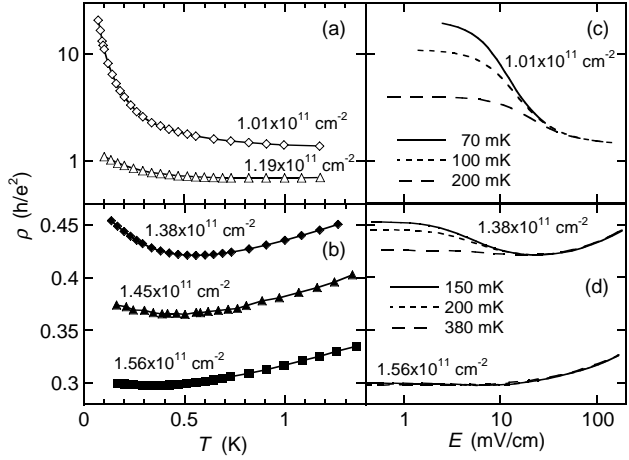}
\caption{(a)-(b) Resistivity as a function of temperature (for $E\rightarrow0$), at 
different densities for sample S1. (c)-(d) Resistivity as a function of electric field  for various 
densities and lattice temperatures.}
\label{fig1}
\end{figure}

\begin{figure}
\includegraphics*[width=8.6cm]{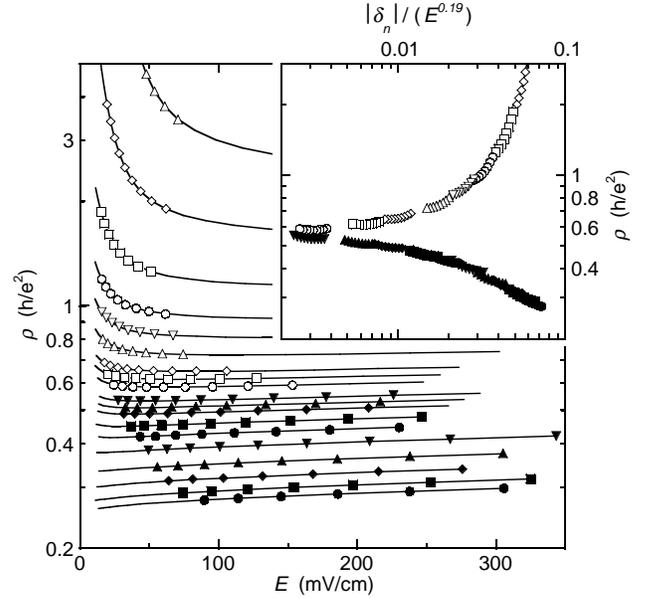}
\caption{Resistivity of sample S1 as a function of electric field at various hole densities: 
from $p_s=0.92\times10^{11}$ cm$^{-2}$ (top curve), 
to $p_s=1.75\times10^{11}$ cm$^{-2}$ (bottom curve). 
The lattice temperature is 120 mK. The lines give 
the whole set of data for $E>10$ mV/cm, while the symbols correspond to the fraction of the data used in 
the curves presented in the inset. 
Inset: resistivity as a function of $\left| \delta_n\right| /E^{0.19}$.}
\label{fig2}
\end{figure}

The MIT features appear in the $\rho(T,p_s)$
dependence
for $E\rightarrow0$ [Fig.\ \ref{fig1}(a)-(b)].
$d\rho/dT<0$ is found for 
$p_s < p_c$$\approx$$1.3\times10^{11}$ cm$^{-2}$ ($r_s\approx 6$), while 
$d\rho/dT>0$ for $p_s>p_c$, at temperatures 
above a threshold which decreases when $p_s$ increases.
At the largest densities $d\rho/dT$ is positive.
The $\rho(T,p_s)$ data 
plotted as a function of $\left| \delta_n \right| /T^b$ 
for $T>0.3-0.5 $ K collapse on two 
branches (not shown), demonstrating a scaling behavior with $b=0.45\pm 0.04$. 
The values 0.35 \cite{Sen1} and 
0.62 \cite{Lam1,Col1} were found in $p$-SiGe.
The MIT characteristics and $E$-field scaling appear also in the 
$\rho(E,p_s)$ curves 
(Fig.\ \ref{fig2}), $p_c$ remaining unchanged.
When $\rho$ is plotted as a function of $\left| \delta_n \right| /E^a$ 
($a$=$0.19\pm 0.02$), 
the curves fall on two branches (inset of Fig.\ \ref{fig2}).
They contain
only a part of the data (symbols on 
Fig.\ \ref{fig2}): removing the low $E$ points is justified 
by $L_E<L_{\Phi}$ (assuming a real QPT), while for 
large $E$, it can be related to a microscopic limit on $L_E$.

\begin{figure}[!t]
\includegraphics*[width=8.6cm]{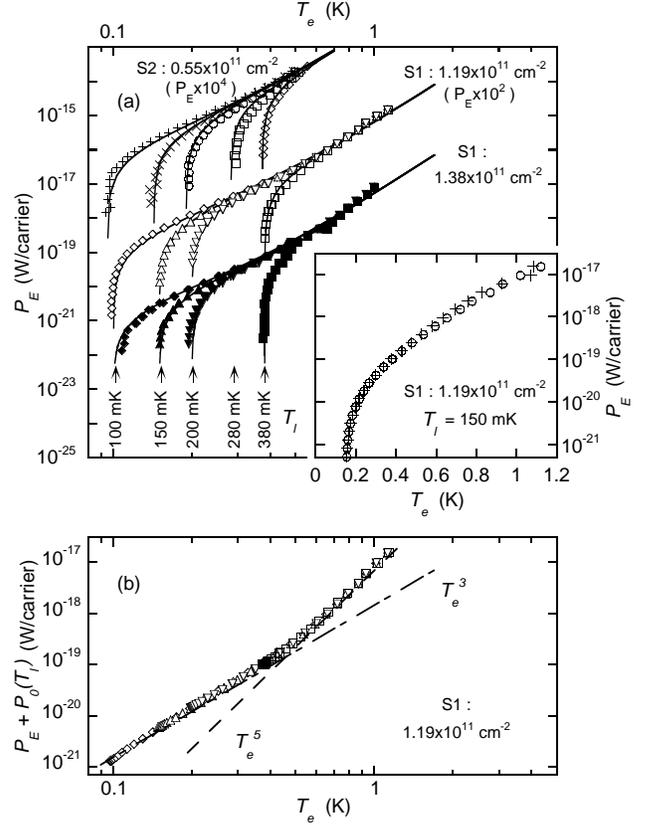}
\caption{(a) Power per hole injected in 
the 2D system as a function of the holes effective temperature, 
for two gated samples and three densities, 
at several lattice temperatures (arrows).
For clarity, the curves corresponding 
to $p_s=1.19\times10^{11}$ cm$^{-2}$ 
(resp. to $p_s=0.55\times10^{11}$ cm$^{-2}$) are shifted 
upwards by 100 (resp. by $10^4$). The lines corresponding 
to the two lower densities are fits to 
$P_E=A(T_e^5-T_l^5)+A'(T_e^3-T_l^3)$; while for the larger 
density a third term $A''(T_e^2-T_l^2)$ has been added 
in the fit formula. Inset: $P_E$ as a function of the effective 
temperature extracted using
the $\rho$ thermometer at $B=0$ ({\smalsymb o}) or 
the damping of the SdH oscillations ($+$).
(b) The sum of the power per hole and a constant $P_0(T_l)$ 
chosen for each $T_l$, as a function of the hole
effective temperature, for various lattice temperatures 
$T_l=100$ mK ($\Diamond$), 150 mK ($\bigtriangleup$), 
200 mK ($\bigtriangledown$), and 380 mK ($\Box$). 
The lines give the slopes of the power laws $T_e^3$ and $T_e^5$.}
\label{fig3}
\end{figure}

Figure\ \ref{fig1} shows that the general shape and the minima 
of $\rho(E)$ are the same as those of $\rho(T)$, suggesting hot hole effects.
Their contribution is studied as follows: 
(i) the $\rho(E)$ dependence is assumed to
result from the hole temperature rise from $T_l$ to $T_e$ 
due to Joule heating; (ii) for 
each value of $E$, 
$\rho$ is used as a 
``thermometer'' giving 
$T_e$ as the temperature $T$ at which the 
same value of $\rho$ is measured for $E\rightarrow0$; 
(iii) the relationship between $T_e$ and 
the power per carrier $P_E=VI/(p_sWL)=E^2/(\rho p_s)$ is used to study 
the validity of the hot carrier assumption (i).
Figure\ \ref{fig3}(a) shows the 
$P_E(T_e)$ dependence for 
various lattice temperatures and for densities 
around $p_c$.

The power loss between 
a degenerate 2D electron system and the lattice 3D acoustic phonons 
is given by \cite{Pri1}
\begin{equation}
P_E=A(T_e^{\alpha}-T_l^{\alpha})+A'(T_e^{\alpha'}-T_l^{\alpha'}).
 \label{Eqchauff1}
\end{equation}
The first term corresponds to the deformation potential coupling, 
with $\alpha=5$ (resp. 7) for weak 
(resp. strong) screening. The second term is piezoelectric coupling, 
to be considered for our 
coherently strained SiGe samples\cite{Xie1,Khi1}, with 
$\alpha'=3$ (resp. 5) for weak (resp. strong) screening.
The exponents may 
be decreased by one in disordered systems because 
of ``dynamic'' rather than ``static'' screening \cite{Cho1}. 
$A$ and $A'$ are prefactors related to the deformation ($\Xi_u$)
and piezoelectric ($e_{pz}$) coupling constants \cite{Pri1,Xie1}.
To test the validity of Eq.\ (\ref{Eqchauff1}) for our measurements, 
we write it:
$P_E+(AT_l^{\alpha}+A'T_l^{\alpha'}\,)=(AT_e^{\alpha}+A'T_e^{\alpha'})$. 
Figure\ \ref{fig3}(b) shows that by adding to $P_E$ a constant
$P_0(T_l)$ chosen separately for each $T_l$, 
the whole set of curves for a given density
falls on the same master curve. Hence 
$P_E(T_e,T_l)+P_0(T_l)$ depends only on $T_e$,
and this dependence is the sum of two power laws. 
We have verified that the $P_0(T_l)$ 
dependence is {\it the same},
thus proving that our data are in agreement with Eq.\ (\ref{Eqchauff1}).
Similar results are obtained for all densities and samples.
For $p_s$ close to $p_c$, a power law with $\alpha'\approx3$ 
(resp. $\alpha\approx5-6$) dominates at low (resp. large) $T_e$. 
These power laws are confirmed 
by fitting the $P_E(T_e)$ data with Eq.\ (\ref{Eqchauff1}).
Fig.\ \ref{fig3}(a) shows the good quality of the fit when
$\alpha = 5$ and $\alpha' = 3$ are imposed.
For $p_s > 1.35\times10^{11}$ cm$^{-2}$,  
a term $A''(T_e^2-T_l^2)$ is added in the fit, corresponding to 
the cooling through the contacts
which increases when $\rho$ decreases\cite{Zie1}. 
$A''$ has the same order of magnitude as the value given by the 
Wiedeman-Franz law. % MODIF
Our data are thus compatible with Eq.\ (\ref{Eqchauff1}), 
the two main cooling processes being 
hole-phonon coupling through weakly screened deformation 
and piezoelectric potentials.
Weak screening at low temperatures has been pointed out 
in {\it p}-SiGe \cite{Xie1,Ans1} and Si-MOSFETs \cite{Zie1,Fle1}. 
$A$ and $A'$ are obtained from a fit with Eq.\ (\ref{Eqchauff1}) assuming
$\alpha=5$ and $\alpha'=3$ [Fig.\ \ref{fig3}(a)]. 
Following Ref.\ \cite{Xie1} we then obtain $\Xi_u=2.7\pm0.3$ eV and 
$e_{pz}=(3.4\pm0.9)\times10^{-3}$ C/m$^2$ 
for all the densities. Our
$\Xi_u$ value is compatible with the 
$\Xi_u\approx 3.0$ eV found in {\it p}-SiGe at 
$p_s=(3.5-13)\times10^{11}$ cm$^{-2}$ \cite{Xie1,Ans1}. 
Our $e_{pz}$ is somewhat
lower than the $e_{pz}\approx1.6\times10^{-2}$ C/m$^2$ 
obtained in {\it p}-Si$_{0.8}$Ge$_{0.2}$
\cite{Xie1}, however in Ref.\ \cite{Ans1} 
$e_{pz}$ was found to be much smaller than in Ref.\ \cite{Xie1}.

In order to further prove the validity of the heating analysis, 
the $T_e$ extracted at $B=0$ is compared to the 
temperature obtained using the damping 
of the SdH oscillations\cite{Xie1,All1,Zie1,Fle2}. 
To get rid of the $\rho_{xx}(T)$ dependence due to carrier 
scattering \cite{Fle2}, the ``thermometer''
is the difference between a minimum of 
$\rho_{xx}(B)$ and the previous maximum.
Its $T$ dependence results from a different physical situation
(the density of states oscillations)
than at $B = 0$. 
The magnetic field values ($B\approx1$ T)
are low to ensure that 
the electron-phonon coupling laws extracted are close to those 
at $B=0$ \cite{Fle2}. As shown in the inset of 
Fig.\ \ref{fig3}(a) for $p_s=1.19\times10^{11}$ cm$^{-2}$ 
and $T_l=150$ mK, the two methods are in very good agreement. 
A similar agreement is obtained for other $T_l$ values and 
for $p_s=3.9\times10^{11}$ cm$^{-2}$.
Thus, the $\rho(E)$ dependence can be attributed to hole heating, 
implying that it does not bring new
physical information on the possible MIT when compared to the $\rho(T)$
dependence.

\begin{figure}
\includegraphics*[width=8.6cm]{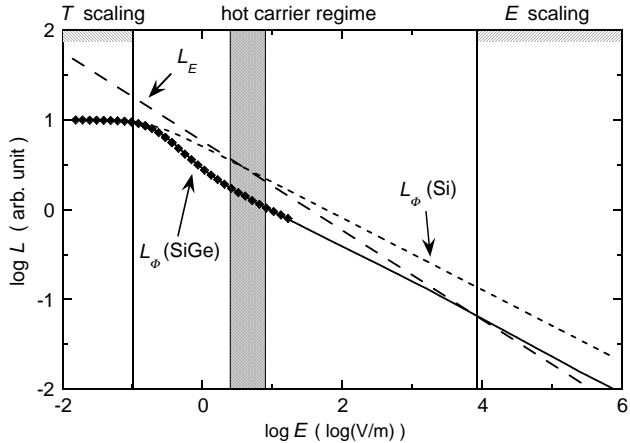}
\caption{The logarithm 
of $L_{\Phi}$ and 
$L_E$ as a function of the electric field. 
The full line represents $L_{\Phi}[T_e(E)]$ calculated 
using Eq.\ (\ref{Eqchauff1}) at
$p_s=1.31\times10^{11}$ cm$^{-2}$ and $T_l=100$ mK. 
The symbols correspond to our experimental points. 
Short-dashed line: $L_{\Phi}[T_e(E)]$
estimated for Si-MOSFETs close to the MIT. Long-dashed line: 
$L_E(E)$ assumed to be the same for SiGe 
heterostructures and Si-MOSFETs. 
The vertical lines separate
the hot carrier region from 
the $T$ and $E$ scaling regions for {\it p}-SiGe. 
The grey area gives the limits above which the 
$E$-field scaling is improved in Ref. [17]}
\label{fig4}
\end{figure}

To study $E$-field scaling in spite of heating effects, 
the condition $L_E(E)<L_{\Phi}[T_e(E)]$ must be fulfilled.
Figure\ \ref{fig4} shows how $L_{\Phi}$ and $L_E$ depend on $E$. The vertical 
scale is arbitrary because $L_{\Phi}$ 
and $L_E$ are not given by the QPT theory, but 
assuming $z=1$ for strongly interacting particles\cite{Son1}, 
$L_{\Phi}\sim T^{-1}$ and $L_E\sim E^{-1/2}$.
Although the ``metallic'' $\rho(T)$ in our samples is compatible with
a Fermi liquid description \cite{Sen2}, it is instructive
to consider the possibility of a QPT:
the full line corresponds to $L_{\Phi}[T_e(E)]$ in our case, 
for $p_s\approx p_c$, 
using the experimental law $T_e(E)$, obtained in the $E$ interval
indicated by the symbols.
An upper limit for $E$ arises since the 
longitudinal potential drop $V=EL$ has to be kept small 
compared to the gate voltage.
As we attributed the measured $\rho(E)$ dependence 
to heating, the $L_E(E)$ line (long dashed) crosses the 
$L_{\Phi}[T_e(E)]$ curve beyond this interval. 
The $E$-field scaling can be investigated only
beyond this crossing point, 
(``$E$ scaling'' interval in Fig.\ \ref{fig4}), provided $L_{\Phi}$
and $L_E$ do not reach their microscopic limit.

The short dashed line in Fig.\ \ref{fig4} is
$L_{\Phi}[T_e(E)]$ for Si-MOSFETs, estimated
using recent measurements of the power loss\ \cite{Zie1,Zie2}.
Again, the $L_E(E)<L_{\Phi}[T_e(E)]$ prescription leads
to a lower limit for $E$. This agrees with the
results of Ref.\ \cite{Lil1} where
the quality of the $E$-field scaling is improved when $E$ is larger
than a minimum value 
(the grey area in Fig.\ \ref{fig4},
corresponds to their cut-offs of 50 to 500 pW). 
For $p$-GaAs, the 
$\alpha=5$ exponent quoted in Ref. \cite{Ken1} leads to a similar situation, 
but the thermal coupling should be larger than in Si-MOSFETs. Unscreened
piezoelectric coupling would lead to an upper
$E$-field limit.
A better knowledge of the power loss laws
would allow a quantitative use of the $L_E(E)<L_{\Phi}(T_e)$
prescription. The field $E_c$ defined by $L_E(E_c)=L_{\Phi}[T_e(E_c)]$
could be extracted as the limit beyond which the experimental $P_E(T_e)$ law 
differs from the power loss law, 
thus yielding
the important physical result
$L_{\Phi}(T)/L_E(E)=[T_e(E_c)/T]^{1/z}(E/E_c)^{1/(z+1)}$.

The ratio of the exponents in
the experimental scaling laws
$\rho(\left| \delta_n \right| /T^b)$ and 
$\rho(\left| \delta_n \right| /E^a)$ has been 
proposed as an indicator of the $\rho(E)$ dependence origin\cite{Son1}.
For the QPT scaling, $a/b=z/(z+1)=0.5$. For carrier heating, 
Eq.\ (\ref{Eqchauff1}) gives
$T_e\sim (E^{2/\alpha})\rho^{-1/\alpha}$, thus $a/b=2/\alpha$
neglecting $A'$, $T_l$ and the $\rho(T)$ dependence. 
We find experimentally $a/b=0.42\pm 0.08$, while
0.4 and 0.67 are expected for our $\alpha$ and $\alpha'$,
thus the $a/b$ criterion can hardly be used\ \cite{SiM1}.

In summary, we have shown that
in a 2D hole system exhibiting the $B=0$ MIT characteristics, with
$T$ and $E$ scaling, the $\rho(E)$ dependence close to the MIT
could be interpreted as being due to
hole heating. Thus,
in our case, the experimental $\rho(E,p_s)$ and
$E$-field scaling are not an indication that the MIT features can
be attributed to a QPT, and
the separate extraction of $\nu$ and $z$
is not possible.
However, there is an $E$ interval defined
by $L_E(E)<L_{\Phi}[T_e(E)]$ where $E$-field scaling
can be investigated independent of hot carrier effects.

The authors thank T. Jolicoeur
for helpful discussions.


\begin{references}
\bibitem{Zie2} The authors of Ref.\ \cite{Zie1} find
$P_E\approx1.5\times10^6n_s^{-3/2}(T_e^5 - T_l^5)$ ($P_E$ in W, $n_s$ in m$^{-2}$),
rather close to the result of Ref.\ \cite{Fle1}. 
We used this equation with $n_s = 10^{15}$ m$^{-2}$. For weak screening
the conductivity dependence of the 
prefactor given in Ref.\ \cite{Cho1} should not be relevant.
\bibitem{SiM1}
Similarly, in Si-MOSFETs, unscreened electron-phonon coupling\ \cite{Zie1,Fle1}
would lead to $2/\alpha$ = 0.4.
\end{references}
\end{document}